\documentclass[twocolumn,pre,showpacs,byrevtex]{revtex4-1}
\usepackage{epsf,graphicx,amssymb}
\usepackage[usenames]{color}

\usepackage{amsfonts}
\usepackage{amsmath}

\begin{document}
\title{Decay of capillary wave turbulence}
\author{Luc Deike}
\author{Michael Berhanu}
\author{Eric Falcon}
\affiliation{Univ Paris Diderot, Sorbonne Paris Cit\'e, MSC, UMR 7057 CNRS, F-75 013 Paris, France, EU}

\date{\today}

\begin{abstract}  
We report on the observation of freely decaying capillary wave turbulence on the surface of a fluid. The capillary wave turbulence spectrum decay is found to be self-similar in time with the same power law exponent than the one found in the stationary regime, in agreement with weak turbulence predictions. The amplitude of all Fourier modes are found to decrease exponentially with time at the same damping rate. The longest wavelengths involved in the system are shown to be damped by viscous surface boundary layer. These long waves play the role of an energy source during the decay that sustains nonlinear interactions to keep capillary waves in a wave turbulent state.   
\end{abstract}

\pacs{47.35.-i, 05.45.-a, 47.52.+j, 47.27.-i}
\maketitle

\section{Introduction}
Waves on ocean surface are the most common example of wave turbulence. Wave turbulence studies the statistical properties of a set of interacting waves where energy is transferred by nonlinear interactions from the forcing scales down to small scales at which energy is dissipated. A statistical theory of wave turbulence was developed in the 60's, the so-called weak turbulence theory which exhibits such an energy transfer in out-of-equilibrium situations \cite{ZakharovBook,Nazarenko}. This theory has been applied to almost every context involving nonlinear waves: astrophysical plasmas, surface or internal waves in oceanography, Rossby waves in the atmosphere, spin waves in magnetic materials, Kelvin wave in superfluid turbulence, nonlinear optics, etc. This theory is based on hypothesis such as weakly nonlinear waves, infinite systems and scale separations between energy source and sink, which may limit its applicability to real systems. In the last decade, the stationary regime of wave turbulence has been studied in laboratory experiments (see \cite{FalconReview} for a recent review), and have shown that the validity domain of this theory can be questionable. Non-stationary wave turbulence, i.e. once the forcing is stopped from a stationary state of wave turbulence, has been much less study although it can provide crucial informations to understand how energy is redistributed between modes during transients (e.g. during a sudden change of wind for ocean waves), and how stationary regime of wave turbulence is reached.

Numerous studies have been performed on non-stationary turbulent systems such as three-dimensional turbulence in wind tunnel \cite{ComteBellot}, in rotating tanks \cite{Moisy06}, and two-dimensional turbulence in: plasma \cite{Bettega09}, soap films \cite{Martin98} and in shallow water \cite{Shats10}, as well as in quantum turbulence \cite{Skrbek12}. For non-stationary wave turbulence, theoretical \cite{ZakharovBook,Connaughton2010} and numerical  \cite{Bigot08,Onorato02} efforts have been performed, whereas experiments are scarce and only concern capillary wave turbulence \cite{Kolmakov04}, or elastic wave turbulence on a thin plate \cite{Miquel11}. For elastic waves, better agreement with weak turbulence theory has been obtained compared to the stationary case where the forcing induces anisotropy \cite{Miquel11}. For the decay of capillary wave turbulence, theoretical works have predicted a self-similar solution in time for the energy spectrum \cite{Falkovitch1995,Kolmakov06} that is compatible with the observations on decaying capillary wave turbulence on the surface of liquid hydrogen \cite{Kolmakov04}. 

In this paper, we report on an experimental study of freely decaying gravity-capillary wave turbulence on the surface of various fluids. The frequency spectrum of the capillary wave amplitude is found to be self-similar during most of the decay. The spectrum exponent of this non-stationary regime is found to be close to the predicted exponent of the stationary regime. Moreover, the energy within the system and all the Fourier modes are found to decay exponentially with time. This is unusual in decaying turbulent systems where energy is supposed to decay as a power law in time \cite{ZakharovBook,Falkovitch1995,Pope}. The longest waves present in the system are the most energetic ones and are shown to be damped by viscosity on the surface, with the inextensible film condition \cite{Lamb}. However, these latter keep enough energy during the decay to sustain capillary waves in a wave turbulent regime without gravity wave turbulence. The amplitude of the capillary spectrum and its cut-off frequency are found to decrease as two different power laws of the total energy contained in the system. Finally, freely decaying capillary wave turbulence can be seen as a capillary wave turbulence in a stationary regime but with a decreasing total energy and a decreasing inertial range with time.

The paper is organized as follows. In Sect. \ref{expset}, the experimental setup and protocol are described. The experimental results are shown in Sect. \ref{dwt}. The wave height power spectrum is described and its properties are analyzed. The evolution of the Fourier modes and of the cut-off frequency of the spectrum are presented during the decay. The origin of wave dissipation is discussed in Sect. \ref{dissipation} whereas Sect. \ref{conclusion} gives the conclusions. Finally, additional experiments in the stationary case are presented in Appendix in order to compare stationary and non- stationary regimes.

\section{Experimental setup and protocol\label{expset}}
The experimental setup is close to the one used in~\cite{Falcon07a}. It consists of a circular plastic vessel, 22 cm in diameter, filled either with mercury or water to a height $h=25$ mm. Water and mercury are used as the working fluid to study the role of the kinematic viscosity on decaying wave turbulence. The properties of mercury (resp. water) are: density, $\rho = 13.6\times10^{3}$ kg/m$^3$ (resp. $10^{3}$ kg/m$^3$), kinematic viscosity $\nu=10^{-7}$ m$^2$/s (resp. $10^{-6}$ m$^2$/s), and surface tension $\gamma=0.4$~N/m (resp. 0.07 N/m). Surface waves are generated by a rectangular plunging wave maker (13~cm in length and 3.5 cm in height) driven by an electromagnetic vibration exciter driven by a random noise (in amplitude and frequency) band-pass filtered typically between 0.1 and 5 Hz. The amplitude of the surface waves at a given location is measured by a capacitive wire gauge plunging perpendicularly to the fluid at rest~\cite{Falcon07a}. The eigenvalues of a circular vessel of radius $R$ are given in \cite{Lamb}, in terms of wave-numbers $k$. By using the dispersion relation for gravity capillary waves $\omega^{2}=(gk+\frac{\gamma}{\rho}k^{3})\tanh{kh}$ ($g$ being the acceleration of gravity, and $\omega$ the angular frequency), the computed eigenvalue frequencies of a circular vessel, corresponding to a wavelength close to $R=11$ cm, are respectively $f^a_{R}=3.2$ Hz for anti-symmetrical modes, and $f^s_{R}=3.9$ Hz for symmetrical ones, whatever the working fluid.

To study freely decaying wave turbulence, a typical experiment is as follows. First, surface waves are generated during a long enough time ($\gtrsim 30$ s) to reach a stationary wave turbulence state. The forcing is then stopped at $t=0$, and the wave amplitude is recorded during a time $T$ long enough to observe the wave damping up to a still state. $T$ is chosen equal to 90 s for mercury and 40 s for water. The experiment then is automatically  iterated $N$ times to generate statistics, and results are averaged (ensemble average denoted by $\langle \cdot \rangle$). We have chosen $N$ large enough to get convergence of the statistics ($N=180$ for mercury and $N=300$ for water). To analyze the different steps of the decay of the wave amplitude signal, the signal is considered on short-time intervals $[t,t+\delta t]$ with $\delta t= 3.5$ s for mercury and 2 s for water (temporal average denoted by $\overline{ \cdot }$). Error bars in the following results have been computed using various values of $\delta t \ll T$ with $1 \leq \delta t \leq 5$ s, and considering different sets of experiments in water or mercury. Such a protocol allows us to study the statistical properties of freely decaying capillary wave turbulence with good accuracy.

\section{Non-stationary wave turbulence \label{dwt}}
\subsection{Temporal decay of the wave amplitude}
\begin{figure}[t]
\begin{center}
\includegraphics[scale=0.5]{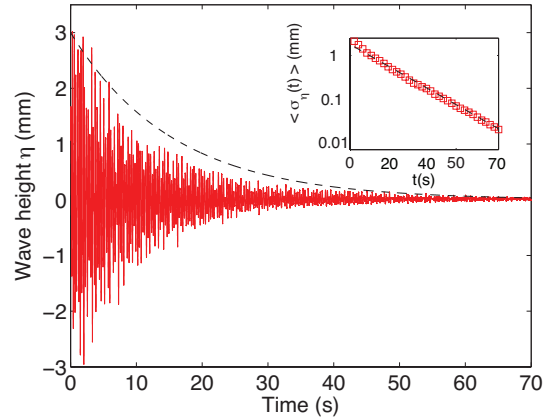}
\caption{(Color online) Decay of the wave amplitude, $\eta(t)$, as a function of time (solid line). Random forcing is working for $t<0$, and is stopped at $t=0$. Mercury. Dashed line corresponds to an exponential fit $\eta (t)=|\eta_{0}|e^{-t/\tau}$ with $\tau=15.5$ s and $|\eta_{0}|=3$ mm. Inset: ($\square$)  temporal decay of the averaged standard deviation of wave amplitude, $\langle \sigma_{\eta} \rangle$, and ($--$) exponential fit $\langle \sigma_{\eta}(t) \rangle=\langle \sigma_{\eta}(0) \rangle e^{-t/\tau}$ with $\tau=15.5$ s and $\langle \sigma_{\eta}(0) \rangle=1$ mm.}
\label{decay}
\end{center}
\end{figure}

Figure \ref{decay} shows the decay of the wave amplitude $\eta(t)$ on the surface of mercury as a function of time for one single realization once the forcing is stopped at $t=0$. The envelope of the wave amplitude, $\eta(t)$ is roughly fitted by $\eta(t)=|\eta_{0}|\exp(-t/\tau)$ (see dashed line) where $\tau$ and $|\eta_{0}|$ are two fitting parameters corresponding to the damping time, and to the absolute value of $\eta$ at $t=0$, i.e. when the forcing is stopped. Moreover, the wave amplitude during its decay exhibits strong fluctuations around its zero mean value before reaching its stationary still state. Inset of Fig.~\ref{decay} shows the temporal evolution of the averaged rms value of the wave amplitude $\langle \sigma_{\eta}(t)\rangle\equiv\sqrt{\langle\overline{\eta^{2}}\rangle}$ computed during each short-time intervals $[t,t+\delta t]$ and then averaged over $N$ experiments. This rms value of the wave amplitude is found to decay exponentially in time with the same damping time $\tau$ as above, and is well fitted by $\langle \sigma_{\eta}(t) \rangle=\langle \sigma_{\eta}(0)\rangle e^{-t/\tau}$ (see dashed line in the inset). Note that similar results are found for water, $\tau$ being the characteristic damping time for a given fluid ($\tau =5$ s for water and $\tau =15.5$ s for mercury). The variance $\langle \sigma_{\eta}(t) \rangle^2$ will be considered as an estimate of the total energy within the system. Indeed, energy of gravity waves $\sim \eta^{2}$, and is greater than capillary energy. As a consequence, the total energy is found to decay as $e^{-2t/\tau}$ with a $\tau/2$ damping time.

\subsection{Time-frequency decay of the spectrum}
\begin{figure}[t]
\begin{center}
 \includegraphics[scale=0.5]{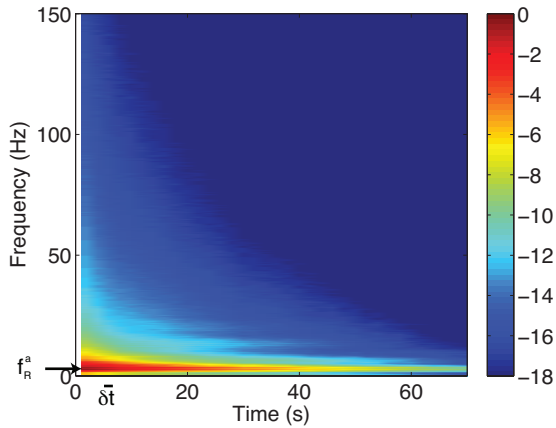} 
\caption{(Color online) Averaged spectrogram $\langle S_{\eta}(f,t) \rangle$ of the wave amplitude during the decay as a function of frequency and time. Color bar is a logarithmic scale of the spectrum amplitude. Pixelization in time corresponds to the short time interval $\delta t=3.5$ s. $f^a_R$ is an eigenvalue of the vessel (see text) remnant of the forcing frequencies (0.1 - 6 Hz).}
\label{spectro}
\end{center}
\end{figure}

To study the decay of the wave amplitude simultaneously in time and in frequency, we use a spectrogram analysis, a well-known MATLAB signal processing tool. Using short-time Fourier transform, the spectrum of the wave amplitude at a time $t^{*}$ of the decay is computed as $S_{\eta}(f,t^{*}) \equiv \int^{t^{*}+\delta t}_{t^{*}} \eta(t)\eta(t+t')e^{-i2\pi ft}dt$ over a short time interval, $[t^{*},t^{*}+\delta t]$. Iterating in time all along the decay, we get the full time-frequency spectrum $S_{\eta}(f,t)$ or spectrogram as a function of time and frequency. This also gives the temporal evolution of the spectrum of one Fourier mode at frequency $f^{*}$: $S_{\eta}(f^{*},t)$. The spectrum is then averaged over $N$ experiments, $\langle S_{\eta}(f,t) \rangle$, and is shown in Fig. \ref{spectro}. A vertical section in Fig. \ref{spectro} at a time $t^{*}$ thus gives the wave amplitude spectrum as a function of frequency at one moment of the decay $\langle S_{\eta}(f,t^{*}) \rangle$. A horizontal section in Fig. \ref{spectro} at a frequency $f^{*}$ thus gives $\langle S_{\eta}(f^{*},t)\rangle$, the temporal evolution of the spectrogram of one Fourier mode during the decay. In Fig.\ \ref{spectro}, $\delta t=3.5$ s is chosen to probe the spectrum for frequencies greater than $2/\delta t \approx 0.6$ Hz. Figure \ref{spectro} shows that low frequencies are more energetic than high ones all along the decay. It also shows that high frequencies are damped faster than low ones. Finally, the most energetic frequency (see arrow in Fig. \ref{spectro}) is $3.1\pm0.1$ Hz. It corresponds to a fundamental eigenvalue of the vessel, $f^a_R=3.2$ Hz that is within the remnant forcing frequencies.

\begin{figure}[t]
\begin{center}
\includegraphics[scale=0.5]{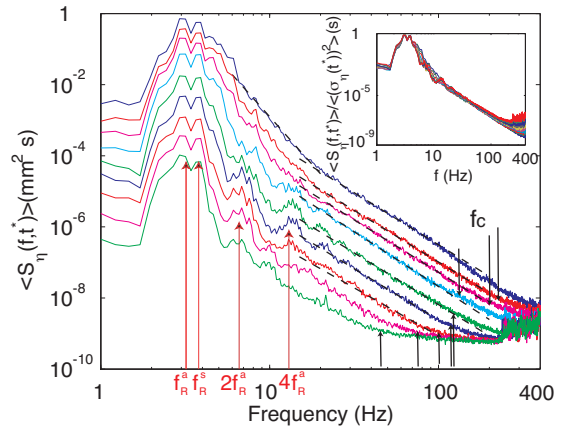} 
\caption{(Color online) Averaged power spectrum $\langle S_{\eta}(f,t^*) \rangle$ at different times $t^*$ of the decay. From top to bottom: $t^*=2$, 5.5, 9, 16, 26.5, 37, 47.5, 58, and 68.5 s. Dashed lines are power laws fits $\sim f^{-5.3}$ for gravity ($6 < f < 25$ Hz - top curve only) and $\sim f^{-2.9}$ for capillary waves ($25 < f < 100$ Hz). Black arrows indicate the cut-off frequencies $f_{c}$ of the capillary wave cascade.  Red (light gray) dashed arrows indicate vessel eigenvalues: $f^a_{R}$ and $f^s_{R}$ remnant of the forcing frequencies (0.1 - 6 Hz), and their harmonics $2f^a_{R}$ and $4f^a_{R}$ (see text). Inset shows the rescaled wave height power spectra $\langle S_{\eta}(f,t^{*}) \rangle / \langle \sigma_{\eta}(t^{*})\rangle ^{2}$.}
\label{spt}
\end{center}
\end{figure}

\subsection{Temporal decay of the power spectrum}
\label{SectSpectrum}
\begin{figure}[t]
\begin{center}
\includegraphics[scale=0.5]{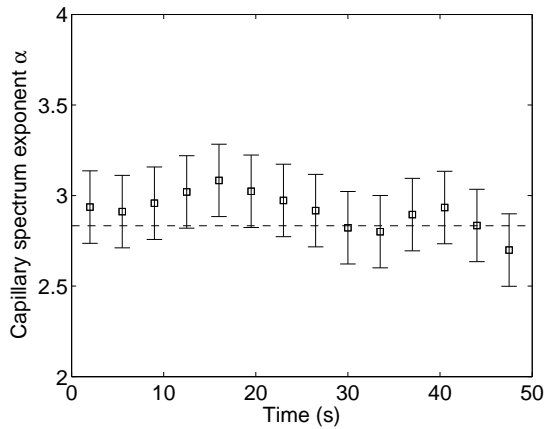} 
\caption{(Color online) Exponent $\alpha$ of the frequency power law spectrum of capillary wave turbulence as a function of time. One has $\alpha=2.9\pm0.2$ during most of the decay ($t<3\tau=45$ s). Dashed line is the theoretical prediction for  stationary capillary wave turbulence \cite{ZakharovBook}.}
\label{exponent}
\end{center}
\end{figure}

Figure \ref{spt} shows the temporal evolution of the power spectrum $\langle S_{\eta}(f,t^{*}) \rangle$ as a function of the frequency at different decay times $t^{*}$. It corresponds to different vertical sections of the spectrogram in Fig. \ref{spectro}, top curve corresponding to the beginning of the decay, just after the forcing is stopped. Time increases from top to bottom, and corresponding curves display the dynamics of the decay in three main phases: i) a remnant of the forcing at the early beginning of the decay, ii) a self-similar decay lasting most of the decay time, and iii) a purely dissipative phase at the end.

i) Remnant of the forcing ($t^*<\tau/3 \simeq 5$ s): the power spectrum just after the forcing is stopped (top curve) is similar to the the one observed in the stationary regime \cite{Falcon07a}. Two frequency-power laws are observed, one corresponding to the gravity wave cascade ($\sim f^{-5.3\pm0.2}$ for $6<f<20$ Hz), and the other to the capillary wave cascade ($\sim f^{-2.9\pm0.2}$ for $20<f<130$ Hz). Crossover between the regimes occurs at $f_{gc}\sim 20$ Hz, while dissipation occurs at higher frequency ($f>130$ Hz).\\

ii) Self-similar phase ($\tau/3< t^*<3\tau \simeq 45$ s): all spectra have the same shape in the capillary regime and exhibit $\langle S_{\eta}(f,t^*) \rangle \sim f^{-\alpha}$ with an exponent $\alpha$ that does not depend on time. Indeed, all frequency-power law fits are parallel in the capillary frequency range. Figure \ref{exponent} shows that the exponent $\alpha$ is roughly independent of time for $0 < t < 3\tau$ with $\alpha=2.9\pm0.2$. This value is in good agreement with the theoretical value $17/6\simeq 2.8$ for the exponent of the stationary capillary wave turbulence \cite{ZakharovBook}. Note that no gravity power law is observed during the self-similar decay of capillary wave turbulence. Moreover, at high frequency, the power law spectrum directly reaches the noise level during the whole self-similar phase. The intercept between the power law and the noise level defines a cut-off frequency of the spectrum and this cut-off is found to decrease continuously during the self-similar decay (see black arrows in Fig. \ref{spt}). Thus the high frequencies vanish first.\\

iii) Purely dissipative phase ($t^*>3\tau$). At long times, no power law is observed in the capillary and gravity frequency ranges. The spectrum has a rounded shape for every frequency.\\

Similar qualitative results are found when water is used as the working fluid. For both fluids (water and mercury), one has the self-similar decay of capillary wave turbulence during  $\tau /3 < t < 3\tau$ with $\tau$ the characteristic damping time of each fluid, whereas the gravity cascade quickly vanishes when $t > \tau/3$. The time $\tau /3$ can be interpreted as the time needed to ``forget" the forcing conditions, or a memory time of the forcing. Although there is a decreasing of the inertial range during the self-similar decay of capillary wave turbulence (see Fig.\  \ref{spt}), all spectra are found to perfectly collapse when rescaled by the variance of the wave amplitude fluctuations, $\langle S_{\eta}(f,t^*)\rangle /\langle \sigma_{\eta}(t^*)\rangle^{2}$, as shown in the inset of Fig. \ref{spt}. The capillary power law cascade is visible up to $3\tau$. This time corresponds to the ``life time'' of the cascade where more than 99\% of the total energy of the system has been dissipated. Indeed, since $\sigma_{\eta}\sim e^{-t/\tau}$, one has $\sigma^2_{\eta}(3\tau)/\sigma^2_{\eta}(0)=e^{-6}$. Thus, we speculate that there is not enough energy within the system to sustain nonlinear interactions and the cascade phenomenology breaks down. Moreover, since the gravity wave cascade vanishes after a time $\tau/3$, we believe that the energy level necessary to sustain gravity wave interactions is much higher than that of capillary waves. Finally, note that two peaks are visible at low frequency in all spectra of Fig. \ref{spt} (3.1$\pm 0.1$ Hz and 3.8$\pm 0.1$ Hz). These frequencies correspond to vessel eigenvalue modes, $f^a_{R}=3.2$ Hz, and $f^s_{R}=3.9$ Hz (\cite{Lamb} and Sec. \ref{expset}), with wavelengths close to the vessel radius $R$. Note also that harmonics of the vessel eigenvalue $2f^a_{R}$ and $4f^a_{R}$ are also observed close to the end of the decay (see Fig. \ref{spt}).

\subsection{Cut-off frequency of the capillary spectrum}
\label{SectCutOff}
\begin{figure}[t]
\begin{center}
\includegraphics[scale=0.5]{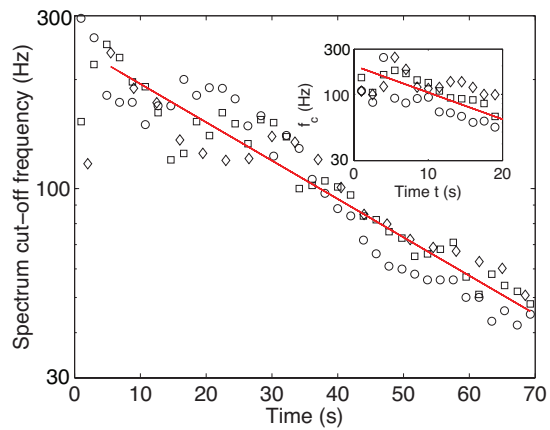}
\caption{(Color online) Cut-off frequency as a function of time. First set of experiments: $\delta t=1.95$ s ($\square$), 3.5 s ($\diamond$). Second set: $\delta t=1.95$ s ($\circ$). Solid line is an exponential fit $f_{c}\sim e^{-at/\tau}$ with $a=0.38$ a fitting parameter and $\tau=15.5$ s the damping time for mercury. Inset: same for water. First set: ($\square$), second set ($\diamond$), and third set ($\circ$). $\delta t=1.5$ s. Solid line is an exponential fit is $f_{c}\sim e^{-at/\tau}$ with $a=0.38$ a fitting parameter and $\tau=5$ s the damping time for water.}
\label{fcut}
\end{center}
\end{figure}
The cut-off frequency $f_{c}$ of the capillary spectrum is shown in Fig. \ref{fcut} as a function of time for different sets of experiments. $f_c$ is defined when the spectrum departs from its power law fit [phase i)] or when it reaches the noise level [phase ii) and iii)] (see black arrows in Fig. \ref{spt}). The cut-off frequency is found to decrease roughly exponentially with time $f_{c}\sim e^{-at/\tau}$ with $a=0.38\pm 0.06$ a fitting parameter, $\tau$ being the characteristic damping time obtained previously for mercury or water. Since $\langle \sigma_{\eta} \rangle \sim e^{-t/\tau}$, one has $f_{c} \sim \langle \sigma^2_{\eta} \rangle^{0.19\pm 0.03}$ during the whole decay for both fluids. Therefore the cut-off frequency is a power law of the amount of energy in the system. This result is compatible with the one obtained in the case of stationary wave turbulence (see Appendix).


\begin{figure}[t]
\begin{center}
\includegraphics[scale=0.5]{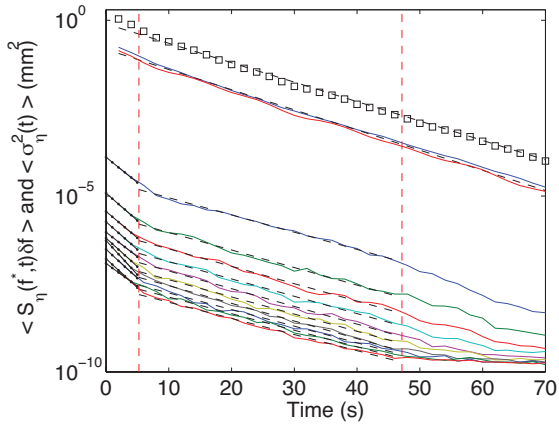} 
\caption{(Color online) Decaying of the total energy $\langle\sigma_{\eta}(t)\rangle^{2}$ ($\square$) and of the Fourier modes $\langle S_{\eta}(f^*,t)\rangle$ as a function of time. Each solid line corresponds to a mode at frequency $f^*$: both top $f^s_{R}=3.2\text{ Hz}$ and $f^a_{R}=3.8$ Hz, from the third top to the bottom $f^*=13$, 23 to 93 Hz with a 10 Hz step. Dashed lines are exponential fits $\sim e^{-2t/\tau}$ for $\tau/3 < t < 3\tau$ (slow damping) and $\sim e^{-5t/\tau}$ for $t < \tau/3$ (fast damping). Vertical dashed lines corresponds to $\tau/3$ and $3\tau$.}
\label{Fourier}
\end{center}
\end{figure}

\subsection{Temporal decay of the Fourier modes}
\label{SectFourier}
Figure \ref{Fourier} shows the temporal evolution of the amplitude of the Fourier modes $\langle S_{\eta}(f^*,t)\rangle$ as a function of time for different $f^*$. It corresponds to different horizontal sections of the spectrogram in Fig. \ref{spectro}. This allows us to compare the total energy in the system $\langle\sigma_{\eta}(t)\rangle^{2}\equiv \langle\overline{\eta^{2}(t)}\rangle=\int_{0}^{\infty}{\langle S_{\eta}(f,t)\rangle df}$ with the amount of energy contained in one single Fourier mode at $f^*$  $\int_{f^*}^{f^*+\delta f}{\langle S_{\eta}(f,t)\rangle df}=\langle S_{\eta}(f^*,t)\rangle \delta f$ with $\delta f=0.24$ Hz being the Fourier transform resolution. First, one observes that the total energy in the system decays exponentially as $\langle \sigma_{\eta}(t) \rangle^{2} \sim e^{-2t/\tau}$ [see $\square$ and the dashed line fit in Fig.\ \ref{Fourier}]. Second, excluding the beginning and the end of the decay, all Fourier modes are found to decay exponentially with the same time scale $\tau/2$ within a 10\% error bar. This is also the damping time of the total energy (see parallel dashed line fits). 
Thus, the damping time is found to be independent of the frequency for interacting waves. This result stands in stark contrast to the one for non interacting waves where the damping time of linear sinusoidal waves is a decreasing function of the wave frequency as a consequence of the viscous damping. 

One might now wonder what feeds the capillary cascade during the decay? The decay of the vessel eigenvalue modes $f^a_{R}=3.2$ Hz and $f^s_{R}=3.8$ Hz are also displayed in Fig.\ \ref{Fourier} (both top solid curves). As for any modes, these fundamental modes both decay exponentially with the same damping time $\tau/2$ as the one of the total energy. Moreover, one also observes that a large amount of energy is contained in these vessel modes. One can estimate that roughly 70\% of the total energy is contained in frequencies between 3 and 4 Hz during all the decay. As a consequence, long waves close to these modes play the role of an energy source that feed the capillary cascade during the decay, and the relaxation of these modes characterize the global decay of the system. The energy decay of these long waves leads to a decrease of the inertial range of the capillary wave turbulence (see Figs. \ref{spt} and \ref{fcut}).

Let us now focus on the decay of Fourier modes at the very beginning and end of the decay. Fig.\ \ref{Fourier} shows that the exponential decay of Fourier modes, within the capillary inertial range, is faster for earlier times, $t < \tau/3$, than for  $\tau/3 < t < 3\tau$, and seems also to be independent of the frequency. The typical time $\sim \tau/5$ of this fast damping could be a measurement of the nonlinear time scale of gravity waves and is found to be smaller than the damping time $\tau/2$. The fast damping is also related to the remnant of the forcing up to a time $\tau/3$ for which 50\% of the total energy that has been dissipated. After this time, the gravity power law spectrum is no longer observed, meaning that there is not enough energy to sustain the gravity wave cascade anymore. The capillary wave turbulence cascade is observed for longer because less energy is needed, and this energy can come from the slow relaxation of the long waves. For $t > 3\tau$, less than 1 \% of the energy remains in the system, and the capillary wave turbulence is no longer observed in Fig. \ref{spt}.

Note that when performing similar experiments with more viscous liquids (aqueous solution of glycerol or silicon oil), no wave turbulence regime is observed, the dissipation time is found to depend on frequency (see Sect. \ref{dissipation}), and is larger than the total energy damping time $\tau/2$. As mentioned above, this behavior is expected when non interacting waves are dissipated by viscosity. As a consequence, decaying capillary wave turbulence is not observed within these fluids because the nonlinear interactions are too weak with respect to the dissipation process.



\section{Origin of wave dissipation}
\label{dissipation}

Dissipation of propagating waves in a closed basin has been studied theoretically and experimentally by various authors \cite{Miles,Landau,Lamb}. The wave height decay at a given frequency has been found to be exponential whatever the nature of the viscous dissipation. Let us define $\mathcal{T}$ the theoretical damping time that depends on the frequency and the nature of dissipation. Viscous damping in fluids can have different origins:  bottom boundary layer ($\mathcal{T}_{B}$), side wall boundary layer ($\mathcal{T}_{W}$), and surface dissipation. Two types of viscous dissipation by the surface are generally considered: either due to the fluid viscosity at the surface $\mathcal{T}^{-1}_{\nu} \sim \nu k^2$ \cite{Landau,Lamb}, or due to a surface boundary layer with an inextensible film $\mathcal{T}^{-1}_{S} \sim (\nu \omega)^{1/2}k$ \cite{Miles,Lamb}. The latter comes from the presence of surfactants/contaminants at the interface that leads to an inextensible surface where fluid velocity should be cancelled at the interface. Note that these surface dissipations are incompatible since they corresponds to two different kinematic conditions at the interface. The decay rate for the wave of frequency $f$ is defined by $\delta\equiv 1/(\mathcal{T}2\pi f)$. The theoretical decay rate for the various types of viscous dissipation in a fluid of arbitrary depth $h$ are \cite{Miles,Landau,Lamb}
\begin{align}
\delta_{\nu} & =\frac{\nu k^{2}}{\pi f} \label{eqnu}\\
\delta_{S} & =\left(\frac{\nu}{4\pi f}\right)^{1/2}\frac{k\cosh^{2}{kh}}{\sinh{2kh}} \label{eqS}\\
\delta_{B} & =\left(\frac{\nu}{4\pi f}\right)^{1/2}\frac{k}{\sinh{2kh}} \label{eqB}\\
\delta_{W} & =\left(\frac{\nu}{4\pi f}\right)^{1/2}\frac{1}{2R}\left(\frac{1+(m/kR)}{1-(m/kR)}-\frac{2k}{\sinh{2kh}}\right) \label{eqW}
\end{align}
where $R$ is the size of the circular vessel, and $m=1$ the anti-symmetrical modes and $m=0$ the symmetrical ones.

\begin{figure}[t]
\begin{center}
\includegraphics[scale=0.5]{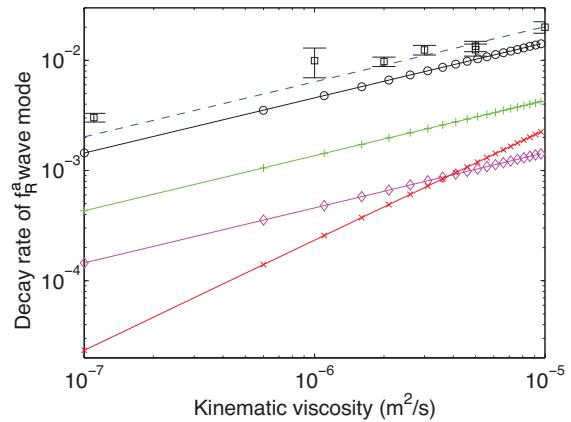} 
\caption{(Color online) Decay rate, $1/(\tau 2\pi f^a_{R})$, of the wave mode at $f^a_{R}=3.2$~Hz as a function of viscosity $\nu$ for various fluids ($\square$). Theoretical decay rate by viscous dissipation: ($\circ$) surface boundary layer, $\delta_{S}$ from Eq. (\ref{eqS}), ($+$) bottom surface layer, $\delta_{B}$ from Eq. (\ref{eqB}), ($\diamond$) wall surface layer,  $\delta_{W}$ from Eq. (\ref{eqW}), and ($\times$) viscous surface $\delta_{\nu}$ from Eq. (\ref{eqnu}). Dashed line is the total theoretical dissipation $\delta_{T}=\delta_{S}+\delta_{B}+\delta_{W}$.}
\label{drate}
\end{center}
\end{figure}

To understand the origin of the wave dissipation in our system, we have performed similar decay experiments with various fluids: water, mercury, silicon oils, and aqueous solutions of glycerol (denoted as x\%GW with x the glycerol percent) to vary kinematic viscosity over two orders of magnitude. Properties of these fluids are listed in Table \ref{tab}. Similar wave relaxation is observed whatever the working fluid: the wave height decays roughly exponentially with time, and so does its rms value $\langle \sigma_{\eta} \rangle$. The most energetic mode $f^a_{R}=$3.2 Hz does not depend on the fluid used as expected, whereas the exponential decay of this mode gives the damping time $\tau$ for each fluid as shown in Table \ref{tab}. Figure \ref{drate} shows the experimental value of decay rate of the frequency $f^a_{R}=$3.2 Hz as a function of the fluid kinematic viscosity ($\square$-symbol), as well as the theoretical decay rates computed for each type of viscous dissipation at the same frequency. Clearly, the experimental decay rate scales as $\nu^{1/2}$ over two decades in viscosity and not as $\nu^1$ as expected by usual viscous dissipation. In our experiment, viscous dissipations by surface boundary layer and bottom boundary layer are the most important while the usual viscous dissipation at the surface is clearly negligible in the damping of the fundamental mode. Friction at the lateral boundary is also negligible. Bottom friction is quite important since the wavelength corresponding to $f^a_{R}$ is close to the vessel radius $R$ (and thus of the order of $2\pi h$); so waves at this frequency are affected by the depth. Finally, the total theoretical dissipation $\delta_{T}=\delta_{S}+\delta_{B}+\delta_{W}$ is found in good agreement with data both qualitatively and quantitatively (see dashed line in Fig.\ \ref{drate}). The fact that viscous dissipation by a surface boundary layer takes over the usual viscous relaxation was previously observed in laboratory experiments with water \cite{Miles,VanDorn,HendersonMiles}, and is also realistic considering waves in a natural context. Indeed, if no particular attention is paid (such as working in clean room, filtered fluid or fluid with low enough surface tension), the surface dissipation by boundary layer dominates the $\nu k^2$ dissipation \cite{HendersonMiles}. Finally, note that the infinite depth condition is satisfied for $f>10$ Hz (i.e. $\lambda <2$ cm and $kh \gg 1$), and thus bottom friction becomes also negligible for capillary waves. Consequently, in our experiments, the dissipation source for capillary waves is only due to to viscous dissipation by a surface boundary layer, and for larger waves  mostly by surface boundary layer since a small amount is also dissipated within the bottom boundary layer.

\begin{table}
\begin{tabular}{|l|c|c|c|c|}
\hline
Fluid & $\rho$ (kg/m$^{3}$) & $\nu$ (m$^{2}$/s) & $\gamma$ (mN/m) & $\tau$ (s)\\
\hline
Mercury & 13 600 & 1.1$\times$10$^{-7}$ & 400 & $15.5\pm1.5$\\
\hline
Water & 1 000 & 10$^{-6}$ & 73 & $5\pm1$\\
\hline
20\% GW & 1 020 & 2$\times$10$^{-6}$ & 70 & $5\pm1$\\
\hline
30\% GW & 1 050 & 3$\times$10$^{-6}$ & 70 & $4\pm0.5$\\
\hline
50\% GW & 1 120 & 5$\times$10$^{-6}$ & 68 & $3.7\pm0.4$\\
\hline
Silicon oil V5 & 1 000 & 5$\times$10$^{-6}$ & 20 & $4\pm0.4$\\
\hline
Silicon oil V10 & 1 000 & 10$^{-5}$ & 20 & $2.5\pm0.4$\\
\hline
\end{tabular}
\caption{Physical fluid properties: density, $\rho$, kinematic viscosity, $\nu$, and surface tension, $\gamma$ \cite{Glycerin}. Corresponding experimental damping time $\tau$ from the temporal decay of the vessel mode of frequency 3.2 Hz.}
\label{tab}
\end{table}

\section{Conclusion}
\label{conclusion}
We report on laboratory experiments on the decay of gravity-capillary wave turbulence on the surface of various fluids. We show that the spectrum of the capillary wave turbulence keeps a self-similar shape during the decay, with a frequency-power law exponent close to the one of the stationary regime. All Fourier modes are found to decay exponentially in time with the same damping rate.  A large amount of the total energy within the system is contained in few longest wave modes (fundamental vessel eigenvalues) which also decay exponentially in time due to viscous dissipation within a surface boundary layer. Since a self-similar capillary wave cascade is observed in time, it means that the typical time of energy flux transfer through scales by nonlinear interactions is small with respect to the natural damping time of the waves. Such a self-similar cascade during the decay of capillary wave turbulence has been predicted theoretically \cite{Falkovitch1995}. However, the total energy is predicted to decay according to a power law in time  \cite{Falkovitch1995}, which is a general result found in various non-stationary turbulent flows \cite{Pope}. 
Our wave turbulence system thus exhibits a unusual behavior where a turbulent cascade regime is observed with a total energy decaying exponentially in time as a result of the viscous decay of the longest wave modes.  It can be interpreted as follows: the energy necessary to sustain the capillary wave turbulence is small with respect to the energy contained in the longest modes that feed the capillary cascade. So even if the major part of the energy is damped by a viscous process, there is still enough energy cascading through the scales to maintain a capillary wave turbulent state provided that the nonlinear time between waves is shorter than the damping time. The amplitude of the capillary spectrum and its cut-off frequency are found to decrease as two different power laws of the energy contained in the system. Both scalings are consistent with those of the stationary regime. This means that freely decaying capillary wave turbulence can be seen as a capillary wave turbulence in a stationary regime but with the total energy and the inertial range decreasing with time.

\begin{figure}[t]
\begin{center}
 \includegraphics[scale=0.5]{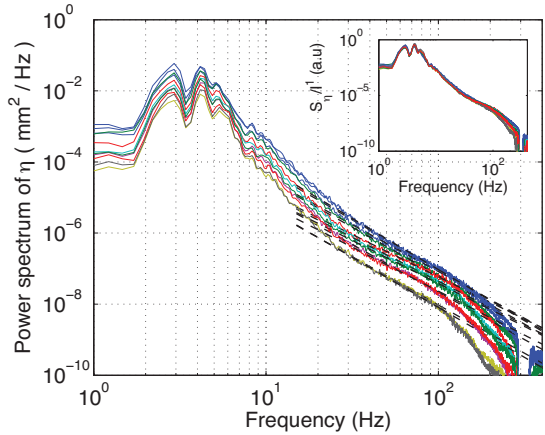} 
\caption{(Color online) Stationary regime. Power spectrum $S_{\eta}(f)$ of wave height for different mean injected powers $\overline{I}$. From bottom to top: $\overline{I}=$ 5, 7, 10, 13, 16, 19, 26 mW. Dashed lines indicate power law fits $\sim f^{-2.8}$. Inset shows the best rescaled spectrum $S_{\eta} / \overline{I}^{\ 1 \pm 0.2}$. Mercury.}
\label{spestat}
\end{center}
\end{figure}
\begin{figure}[t]
\begin{center}
 \includegraphics[scale=0.5]{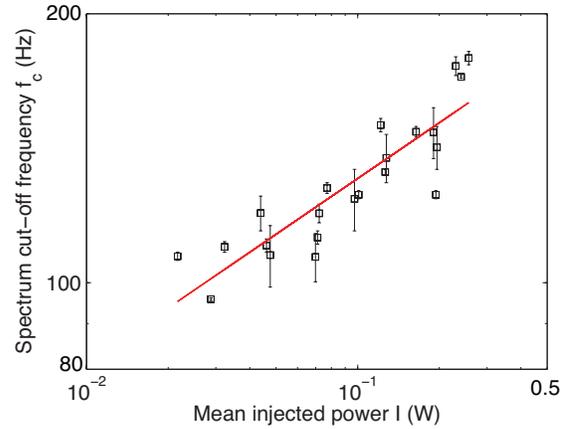} 
\caption{(Color online) Stationary regime. Cut-off frequency $f_{c}$ of the capillary wave spectrum as a function of the mean injected power $\overline{I}$. Solid line is a power law fit $f_{c}\sim \overline{I}^{\ 0.2\pm0.05}$. Mercury.}
\label{cutoffperm}
\end{center}
\end{figure}

\begin{acknowledgments}
We thank C. Laroche for technical help. This work has been supported by ANR Turbonde BLAN07-3-197846.
\end{acknowledgments}

\appendix*
\section{Stationary wave turbulence \label{perm}}
Let us compare the experimental results found in the non-stationary case (see above) with that of the stationary regime of capillary wave turbulence. To do this, additional experiments were performed in a stationary regime with decreasing injected energy. The experimental setup in the stationary case is the same as described in Sect. \ref{expset} (see also \cite{Falcon07a}) except that the wavemaker is continuously driven by a stochastic forcing (both in amplitude and frequency in a range of typically 0.1 to 6 Hz). The wave height $\eta(t)$ is then recorded during 300 s, and its power spectrum is computed over the whole duration of the signal recording. The force $F(t)$ applied by the shaker to the wavemaker and the velocity $V(t)$ of the wavemaker are measured to access to the injected power $I=F \times V$ into the waves \cite{Falcon07a}.

For such a stationary regime, one finds that mean injected power $\overline{I} \sim \sigma_{\eta}^{2}$. Figure \ref{spestat} shows the wave height spectra for different forcing amplitude $5 \leq \overline{I} \leq 30$ mW. Whatever the forcing, two power laws are observed: one in the gravity wave regime, $S_{\eta}\sim f^{-5.2\pm0.2}$; and one in the capillary range of frequency, $S_{\eta}\sim f^{-2.8 \pm 0.2}$ as previously reported \cite{Falcon07a}. The inertial range of the capillary spectrum also increases when $\overline{I}$ is increased. The power law exponent of the capillary regime is found to be independent of $\overline{I}$, and the inset of Fig. \ref{spestat} shows that $S_{\eta}\sim \overline{I}^{\ 1\pm0.2}$ in agreement with previous experimental results \cite{Falcon07a,Xia10}. Since $S_{\eta}\sim \overline{I}f^{-2.8}$ and $\overline{I}\sim\sigma_{\eta}^{2}$, one thus has $S_{\eta}\sim \sigma_{\eta}^{2}f^{-2.8}$ for the stationary regime which is the same result as the one found experimentally for freely decaying capillary turbulence (see Sect.~\ref{SectSpectrum}). 

In the stationary case, the cut-off frequency $f_c$ of the capillary wave turbulence spectrum is defined when the spectrum departs from its power law fit (see Fig. \ \ref{spestat}). Figure \ref{cutoffperm} then shows that $f_{c}$ increases as a power law in $\overline{I}$, i.e.  $f_{c}\sim\overline{I}^{\ 0.2\pm0.05}$ leading to $f_{c}\sim(\sigma_{\eta}^{2})^{0.2\pm0.05}$ which is the same result as the one found in the freely decaying case (see Sect. \ref{SectCutOff}). 

Finally, note that in a stationary regime, the cut-off frequency $f_c$ is generally estimated by balancing the nonlinear interaction scale time $\tau_{NL}$ with the dissipative time scale $\tau_D$:  $\tau_{NL}(f_c) \sim \tau_{D}(f_c)$ \cite{ZakharovBook,Brazhnikov02}. However, here, we have no measurement of $\tau_{NL}$ at capillary scales. On the other hand, the cut-off frequency of the capillary wave turbulence on liquid hydrogen has been reported to scale as $f_{c} \sim \eta_{p}^{4/3}$ with $\eta_{p}$ the wave amplitude at the forcing frequency \cite{Brazhnikov02}. Assuming that $\eta_{p} \sim \sigma_{\eta}$ leads to $f_{c} \sim \sigma_{\eta}^{4/3}$, and thus a $2/3$ exponent for the variance which differs from the $0.2$ exponent found here. Several differences between both experiments can explain this fact. In our case, the stochastic forcing by a wave maker leads to a continuous power law spectrum, whereas the parametric forcing at a single frequency of Ref.\ \cite{Brazhnikov02} leads to a discrete spectrum of peaks of decreasing maximal amplitudes. Moreover, the dissipation nature is different since surface viscous boundary layer is shown to be the main dissipative mechanism for water waves whereas a $\nu k^2$ dissipation is assumed for liquid hydrogen \cite{Brazhnikov02}.

\end{document}